\documentclass{article}

\usepackage[british]{babel}
\usepackage[useregional]{datetime2}
\DTMlangsetup[en-GB]{showdayofmonth=false}
\usepackage[numbers]{natbib}
\usepackage[letterpaper,top=2cm,bottom=2cm,left=3cm,right=3cm,marginparwidth=1.75cm]{geometry}
\usepackage{amsmath}
\usepackage{amsfonts}
\usepackage{amssymb}
\usepackage{mathtools}
\usepackage{wrapfig}
\usepackage{graphicx}
\usepackage{tikz}
\usepackage[colorlinks=true, allcolors=blue]{hyperref}
\usepackage[page]{appendix} %

\usepackage{graphicx}
\usepackage{subcaption}

\renewcommand{\vec}[1]{\boldsymbol{\mathbf{#1}}}
\renewcommand{\v}{\vec}

\usepackage[letterpaper,top=2cm,bottom=2cm,left=3cm,right=3cm,marginparwidth=1.75cm]{geometry}
\usepackage{amsmath}
\usepackage{amssymb}
\usepackage{amsfonts}
\usepackage{cancel}
\usepackage{mathtools}
\usepackage{graphicx}
\usepackage{tikz}
\usetikzlibrary{arrows}
\usepackage[colorlinks=true, allcolors=blue]{hyperref}
\usepackage[page]{appendix} %

\renewcommand{\vec}[1]{\boldsymbol{\mathbf{#1}}}
\renewcommand{\v}{\vec}

\newcommand{\sumiN}{\sum_{i=1}^N}

\usepackage{authblk}

\title{\vspace*{0cm}{\textbf{\Large{Closed-form solutions for generic $N$-token AMM arbitrage}}}}
\author{Matthew Willetts}
\author{Christian Harrington}
\affil{QuantAMM.fi}
\begin{document}
\maketitle

\begin{abstract}
Convex optimisation has provided a mechanism to determine arbitrage trades on automated market markets (AMMs) since almost their inception.
Here we outline generic closed-form solutions for $N$-token geometric mean market maker pool arbitrage, that in simulation (with synthetic and historic data) provide better arbitrage opportunities than convex optimisers and is able to capitalise on those opportunities sooner.
Furthermore, the intrinsic parallelism of the proposed approach (unlike convex optimisation) offers the ability to scale on GPUs, opening up a new approach to AMM modelling by offering an alternative to numerical-solver-based methods.
The lower computational cost of running this new mechanism can also enable on-chain arbitrage bots for multi-asset pools.
\end{abstract}
\section{Introduction}

Automated market makers (AMMs), of which most decentralized exchanges (DEXs) are examples, are some of the most important Decentralized Finance (DeFi) infrastructure this is.
In these systems Liquidity providers deposit assets (LPs) that is used to enable traders to exchange tokens against the deposited capital.
These systems provide close-to-market prices in their trades as there is an economic incentive for them to do so: quoted prices depend inversely on the pool's reserves, meaning that if the quoted price for an asset is too high a trader can gain by extracting that asset out of the pool (exchanging something else in) and vice versa if the quoted price is too low the trader can gain from trading that asset out of the pool.
These trades are \emph{arbitrage trades}, are carried out by the pool's arbitrageurs, and have the effect of bringing the pool's quoted prices closer to the current market prices.

We are interested in modelling the arbitrage opportunities that arise on multi-token (i.e. $N>2$) geometric mean market makers, in particular finding the optimum trade for the arbitrageur to perform to maximise their return at current market prices.
This takes us beyond swaps, as in general the arbitrage opportunity requires the trading of multiple different assets into or out of the pool, or both.

This problem can be attacked numerically using linear programming/convex optimisation~\cite{angeris2021multi}.
Here we take a different approach, deriving closed-form analytical expressions for the optimal arbitrage trade.
Fundamentally, these closed-form expressions for the optimal multi-asset arbitrage trade enable new and fine-grained mathematical analysis of the properties of the trade.
These closed form expressions are particularly useful in the context of Temporal-function Market Making (TFMM)~\cite{tfmm}, where pools' weights change with time, as they enable modelling of the arbitrage trade (and resulting changes in pool reserves) \emph{through which one can take gradients for the purpose of optimisation}.

Unlike a linear program, these closed form expressions require an \emph{a priori} known amount of arithmetic operations to calculate, and for reasonable numbers of assets ($N\lessapprox7$) are substantially quicker to perform than convex optimisation (see Figure~\ref{fig:time_to_calc}).

\section{Optimal Multi-Asset Trades in the presence of fees}

The pool contains $N$ tokens, so a trader has a vector $\v \Delta$ of token amounts being traded to the pool for another set of tokens $\v \Lambda$, where entries in each are $\geq 0$.
Where $\Delta_i>0$, $\Lambda_i=0$: tokens cannot be traded for themselves.\footnote{Equivalently, $\exists N\in\mathbb{N}, N\geq2:~\forall \v\Lambda\in \left(\mathbb{R}^{+}_0\right)^N,\,\v\Delta\in \left(\mathbb{R}^{+}_0\right)^N, \v \Delta\cdot\v\Lambda=0$.}
With fees $(1-\gamma)$, for a trade to be accepted it must be the case that
\begin{equation}
    \prod_{i=1}^N \left(R_i + \gamma \Delta_i - \Lambda_i\right)^{w_i} \geq k.
    \label{eq:tfmm_with_fees}
\end{equation}
where $\v R= \{R_i\}_{i=1,...,N}$, $\v w =\{w_i\}_{i=1,...,N}$, $\forall i,\, 0<w_i<1$, $\sumiN w_i=1$ and  $k=\prod_{i=1}^N R_i^{w_i}$.
While any trade that increases $\prod_{i=1}^N \left(R_i + \gamma \Delta_i - \Lambda_i\right)^{w_i}$ above $k$ is accepted, a trader does better if their trade maintain $k$.\footnote{The inequality trading constraint is of use, however, as it enables convex optimisation (which demanding equality does not), with the obtained solution lying on the level set for which equality is maintained \cite{angeris2021multi}.}
In constructing the optimal arbitrage trade, first we have to know \emph{which} tokens to trade in and \emph{which} to withdraw.
We will return to that question in Section~\ref{sec:no_arb_modelling}.

It is notationally simpler if we combine the inwards and outwards trade legs into one variable $\v \Phi:=\v \Delta - \v \Lambda$.
If token $i$ is not being traded, $\Phi_i=0$. 
We introduce a \emph{trade signature} $\v s$ of length $N$, $s_i\in\{-1,0,1\}$ which encoders whether a token is being extracted from the pool, not being traded, or is being added to the pool.
We can index over the \emph{active} tokens in the pool, those with $s_i\neq0$.
We define the set $\mathcal{A}=\{i\in[N]|s_i\neq0\}$, and we define an active-token normalised set of weights $\breve{w}_i:=w_i/\sum_{j\in\mathcal{A}} w_j$.
We also introduce an auxiliary variable $\v d= \mathbb{I}_{\v s=1}$ for application of fees.\footnote{This turns our problem into a mixed-integer convex program if we have our trading-function invariance imposed via an inequality, c.f. the above footnote.}

We can rewrite Eq~\eqref{eq:tfmm_with_fees} as
\begin{equation}
    \prod_{i\in\mathcal{A}} \left(1 + \frac{\gamma^{d_i} \Phi_i}{R_i}\right)^{\breve{w}_i} = 1,
    \label{eq:tfmm_with_fees_phi}
\end{equation}
a reduced form of the trading function (Appendix~\ref{app:reduced}).
The we can form the multi-asset objective as
\begin{equation}
    \mathcal{L} = -\sum_{i\in\mathcal{A}}\left(m_{p,i} \Phi_i\right) - \lambda \left( \prod_{j\in\mathcal{A}} \left(1 + \frac{\gamma^{d_j} \Phi_j}{R_j}\right)^{\breve{w}_j} - 1\right).
\end{equation}
Taking partial derivatives w.r.t. $\Phi_i$ and setting to $0$
\begin{align}
    \frac{\partial \mathcal{L}}{\partial \Phi_i} = -m_{p,i} - \lambda \frac{\breve{w}_i \left(1 + \frac{\gamma^{d_i} \Phi_i}{R_i}\right)^{\breve{w}_i}\gamma^{d_i} \prod_{\substack{j\neq i \\ j\in\mathcal{A}}} \left(1 + \frac{\gamma^{d_j} \Phi_j}{R_j}\right)^{\breve{w}_j}}{\left(1 + \frac{\gamma^{d_i} \Phi_i}{R_i}\right)}=0&\\
    \Rightarrow -m_{p,i} - \lambda \frac{\breve{w}_i\gamma^{d_i} \prod_{j=1}^N \left(1 + \frac{\gamma^{d_j} \Phi_j}{R_j}\right)^{\breve{w}_j}}{\left(1 + \frac{\gamma^{d_i} \Phi_i}{R_i}\right)}=0&\\
    \Rightarrow \lambda = -\frac{m_{p,i} R_i \left(1 + \frac{\gamma^{d_i} \Phi_i}{R_i}\right)}{\breve{w}_i \gamma^{d_i}}\qquad\qquad\qquad&
\end{align}
This gives us $|\mathcal{A}|$ non-degenerate equations each solving for $\lambda$.
Putting any two into equality,
\begin{align}
    \frac{m_{p,i} R_i \left(1 + \frac{\gamma^{d_i} \Phi_i}{R_i}\right)}{\breve{w}_i \gamma^{d_i}} &= \frac{m_{p,j} R_j \left(1 + \frac{\gamma^{d_j} \Phi_j}{R_j}\right)}{\breve{w}_j\gamma^{d_j}}\\
    \Rightarrow \left(1 + \frac{\gamma^{d_j} \Phi_j}{R_j}\right) &= \frac{\breve{w}_j \gamma^{d_j}}{\breve{w}_i \gamma^{d_i}}\frac{m_{p,i} R_i}{m_{p,j} R_j}\left(1 + \frac{\gamma^{d_i} \Phi_i}{R_i}\right)
    \label{eq:ratio_rearranged}
\end{align}
We can now sub in Eq~\eqref{eq:ratio_rearranged} to Eq~\eqref{eq:tfmm_with_fees_phi} for all $j\neq i$ and rearrange (Appendix~\ref{app:basket_optimal}) to obtain
\begin{equation}
    \forall i \in \mathcal{A},\,\Phi_i=\gamma^{-d_i}\left(\breve{k}\left(\frac{\breve{w}_i \gamma^{d_i}}{m_{p,i}}\right)^{1-\breve{w}_i}\prod_{\substack{j\neq i \\ j\in\mathcal{A}}}\left(\frac{m_{p,j}}{\breve{w}_j \gamma^{d_j}}\right)^{\breve{w}_j}-R_i\right),
\end{equation}
where $\breve{k}=\prod_{i\in\mathcal{A}} R_i^{\breve{w}_i}$
This gives the optimal arbitrage trade for a multi-token/basket trade for geometric mean market makers.

\paragraph{Comment on derivation}
When first looking at these problems of optimal trades, it can seem that closed-form solutions are not possible as we have to impose inequality constraints on the trade (e.g. that $\Phi_i>0$ where $s_i=1$ in Eq~\eqref{eq:tfmm_with_fees_phi}) meaning we have (Karush–Kuhn–Tucker) KKT conditions.

Our derivations works because a) we externally chose the trade signature, b) the signature allows for tokens to be untouched in the trade and c) the solutions to the full problem satisfy complementary slackness.
Together these mean that we do not have to follow a KKT formulation, instead we get closed-form expressions via Lagrange multipliers.

Finally, note that in the case that fees are not present, the modelling task is much simpler~\cite{tfmm}.
Various approaches and results can then be layered on top of the zero-fee-case model (e.g. the bound in \cite{angeris2020does}).
Here we are interested in full treatment of the effect of fees.

\subsection{Finding the optimal trade signature}
To use our results, it seems we need \emph{a priori} the arbitrage trade's signature $\v s$.
There is a simple fix: run through all possible variants of $\v s$ calculating $\v\Phi(\v s)$, and for each $\v\Phi(\v s)$ calculate its return to the arbitrageur $-\sumiN\left(m_{p,i} \Phi_i\right)$ and check that it fulfils the trade invariant Eq~\eqref{eq:tfmm_with_fees_phi}.
The best such trade is then the optimal.
If no checked trade fulfils Eq~\eqref{eq:tfmm_with_fees_phi} while having a return $>0$, the best trade is $\v\Phi=0$ and we are in the no-arb region.

This computationally-intensive parts of this approach (calculating $\v\Phi(\v s)$, $-\sumiN\left(m_{p,i} \Phi_i\right)$ and checking Eq~\eqref{eq:tfmm_with_fees_phi}) are embarrassingly-parallel over different settings of $\v s$.

\begin{wrapfigure}[15]{r}{0.4\textwidth}
\centering
        \centering
        \hspace{-1.5em}
    \includegraphics[width=0.4\textwidth]{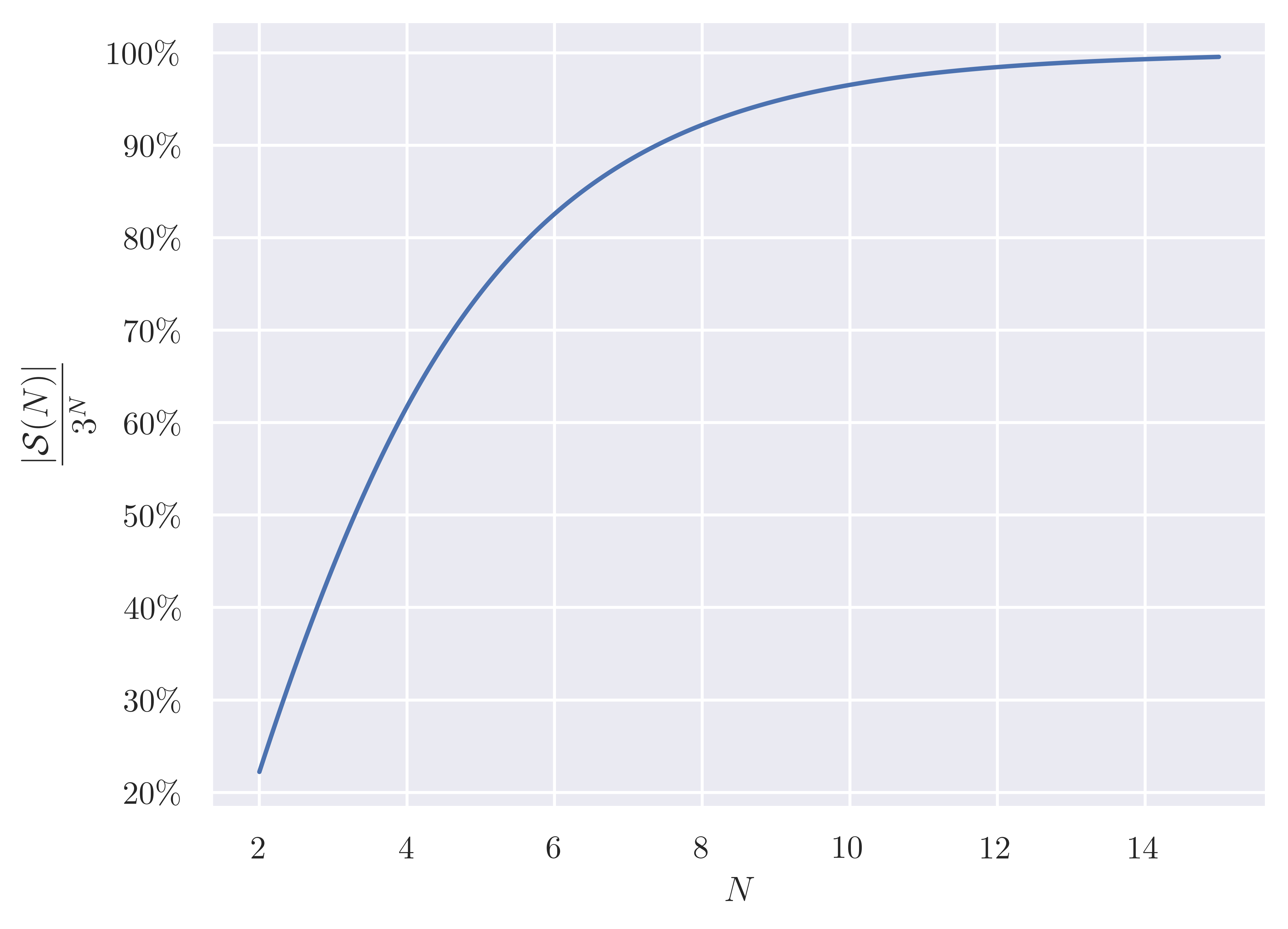}
    \caption{Number of valid trade signatures as a fraction of all possible unconstrained signatures.}
    \label{fig:ratio_n}
\end{wrapfigure}

What are the sets of $\v s$ that one has to check for a given pool size $N$?
We denote set of valid trade signatures $\mathcal{S}(N)$.
The upper limit for $|\mathcal{S}(N)|$, the number of signatures a pool can have, is $3^N$ (each entry $s_i$ can be one of $\{-1,0,1\}$), but not all combinations are valid.
A signature $\v s$ has to have at least one entry of $1$ and at least one $-1$ (at least one token has to be traded for at least one other token).
For pools with $N<6$ or so, these restrictions mean the number of valid signatures is $<80\%$ of the naive $3^N$ calculation.
We plot $|\mathcal{S}(N)|$ as a fraction of $3^N$ in Figure~\ref{fig:ratio_n}.
As $N$ increases, this ratio tends to one.
For reference, $|\mathcal{S}(N=3)|=12$ and $|\mathcal{S}(N=4)|=50$.

Another possibility is create a computationally-less-intensive heuristic we can use to calculate the trade signature/trade direction--though for low $N$-token pool running through all variants is still considerably faster than convex optimisation even prior to any other speedups.
An example of such a heuristic is provided in Appendix~\ref{app:heuristic}.

\section{Simulation results}

Fundamentally, arbitrage trades are how AMM pools' reserves update.
In modelling our results in silico and benchmarking against existing (numerical) approaches we are interested in both the run time of our approach and in how high-fidelity it is in finding arb opportunities.
Both these areas matter whether one is interested in using our approach for simulating AMM pool rebalancing (for example in a backtest) or for using it to find and act on arbitrage opportunities in real time.

\begin{figure}[h]
    \centering
    \includegraphics[width=0.5\textwidth]{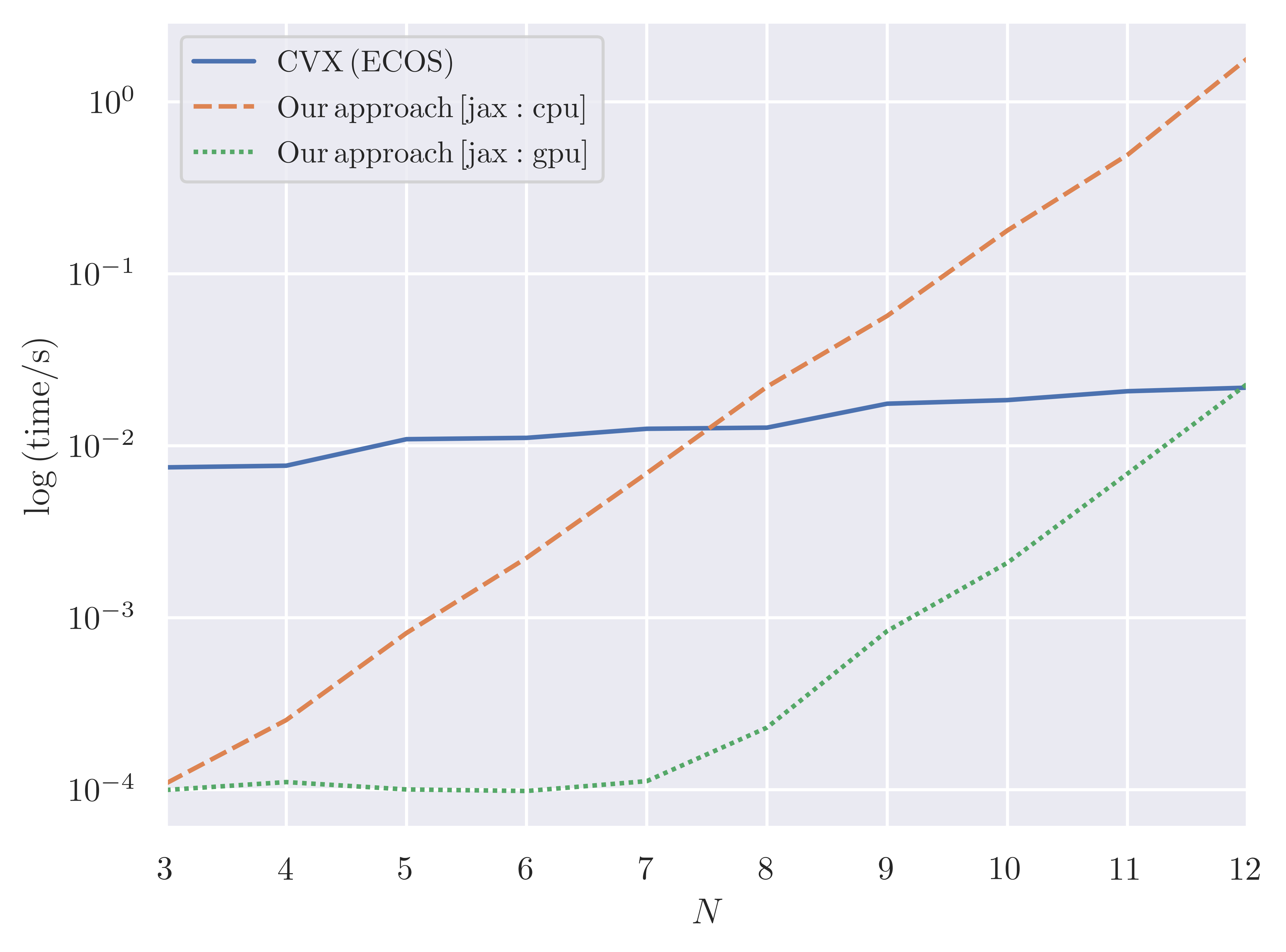}
\caption{Logarithmic plot of the time taken to calculate an optimal trade as a function of number of pool assets $N$. All experiments were run in python, using CVXPY for convex optimisation. We used a MacBook Air M2 for all CPU results and a workstation with a GPU with compute capability 8.9.}    
    \label{fig:time_to_calc}
\end{figure}
\subsection{Run time}

For large values of $N$, standard convex solvers scale better than the closed form solutions, given the naive brute force methodology for finding the signature of the optimal trade requires roughly $3^N$ computations (though new heuristics may remove the advantage).
As demonstrated in Figure~\ref{fig:time_to_calc}, the scalability of the closed form solution can be drastically helped because unlike current convex solutions, it can be parallelised (for both CPUs and GPUs).
While this does only delay the point at which convex solvers become faster, even with a single GPU it delays the point at an $N$-token number that is rarely, if at all, seen in contemporary AMMs.

\subsection{Effectiveness}

We create simulated, independent, potential arbitrage opportunities.
For each we start with a pool initially at equilibrium (pool prices match market prices) and then randomly sample new market prices that deviate slightly, which have a chance of pushing the pool out of the no-arb region.
\begin{equation}
    m_{p,i}\sim\mathrm{Uniform}(0,1),\quad \v R_{\mathrm{initial}} = V_0 \frac{\v w}{\v m_p}, \quad u_i\sim \mathrm{Uniform}(0,1), \quad \v m_p^* = \v m_p + a \v u, \nonumber
\end{equation}
where $V_0$ is the initial value of the pool in the num\'eraire of $\v m_p$, each trial's $\v w$ is chosen to be close to uniform weights $w_i = \frac{1}{N}$ (as convex optimisation is less stable when $\v w$ is close to one-hot) and $\gamma=0.05$.

\begin{figure}[h]
    \centering
    \includegraphics[width=0.5\textwidth]{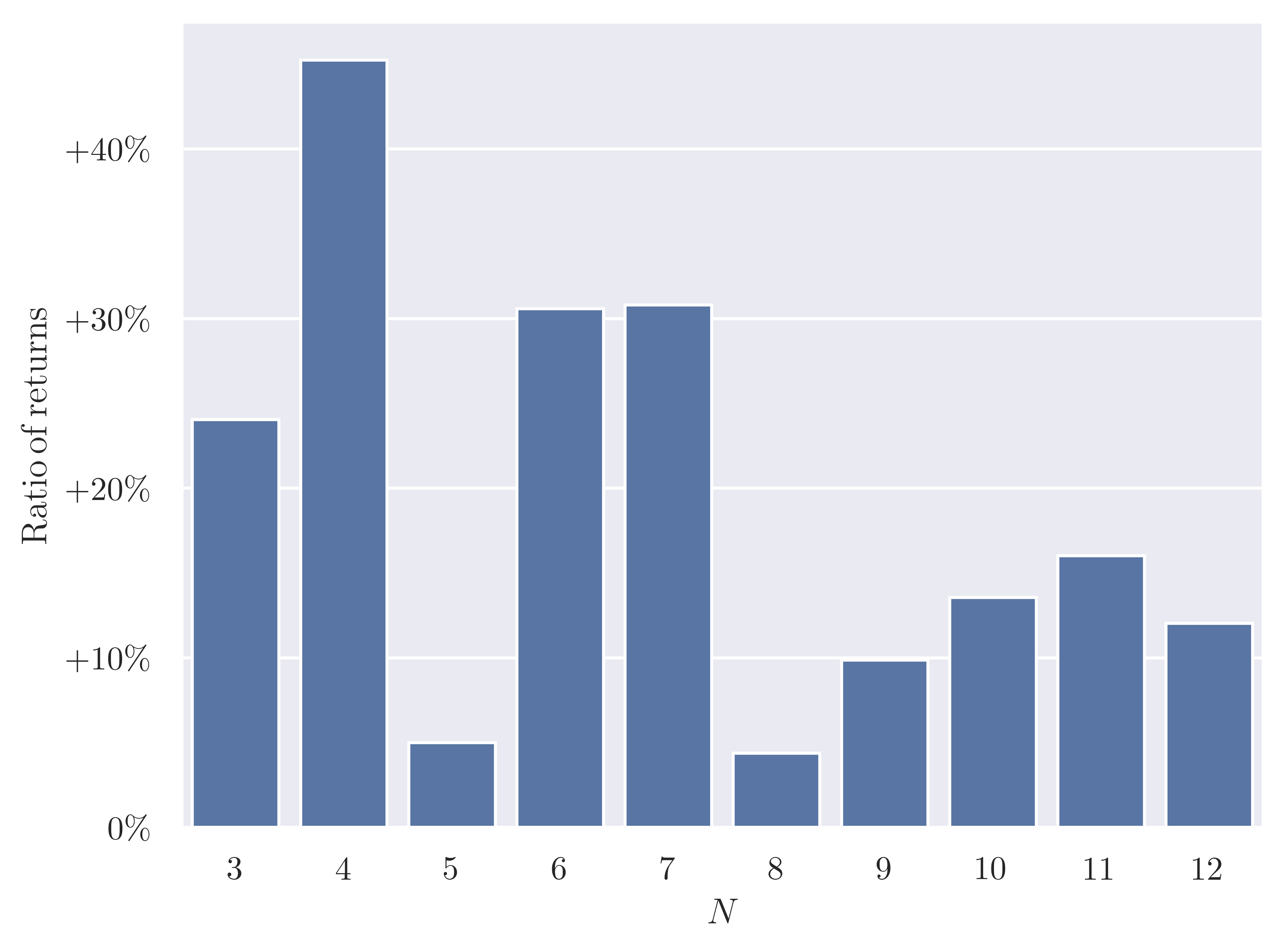}
     \caption{Comparing arbitrage returns from our approach as a function of the number of pool assets against numerical convex optimisation.}
     \label{fig:cvx_diff}
\end{figure}

We perform 120,000 independent random trials, and for each record the result of convex optimisation and our approach. Figure~\ref{fig:cvx_diff} compares the $N$-token closed to convex results.
This demonstrate the superiority of the $N$-token approach across pools to find arbitrage opportunities.

\label{sec:no_arb_modelling}
\subsection{Duelling arbitrageurs}
\label{sec:arb_modelling}

If you run multiple separate simulations of an $N$-token pool over time, in each updating the pool according to the arbitrage trades produced by a particular arbitrage algorithm, the algorithm that is \emph{less good} at constructing arbitrage trades will often show \emph{greater profit}.
Why is this unexpected result found?
In a less-than-ideal system, the arbitrage opportunity is only detected and acted on further away from the no-arb region of the pool.

As there is no competition between arbitrageurs (we are comparing separate runs), the less-good algorithm wins by `farming' the pool, letting the arbitrage opportunity build up and up before eventually being reaped.

\begin{figure}[htbp]
\centering
\begin{subfigure}[b]{0.45\linewidth}
        \centering
    \includegraphics[width=1\textwidth]{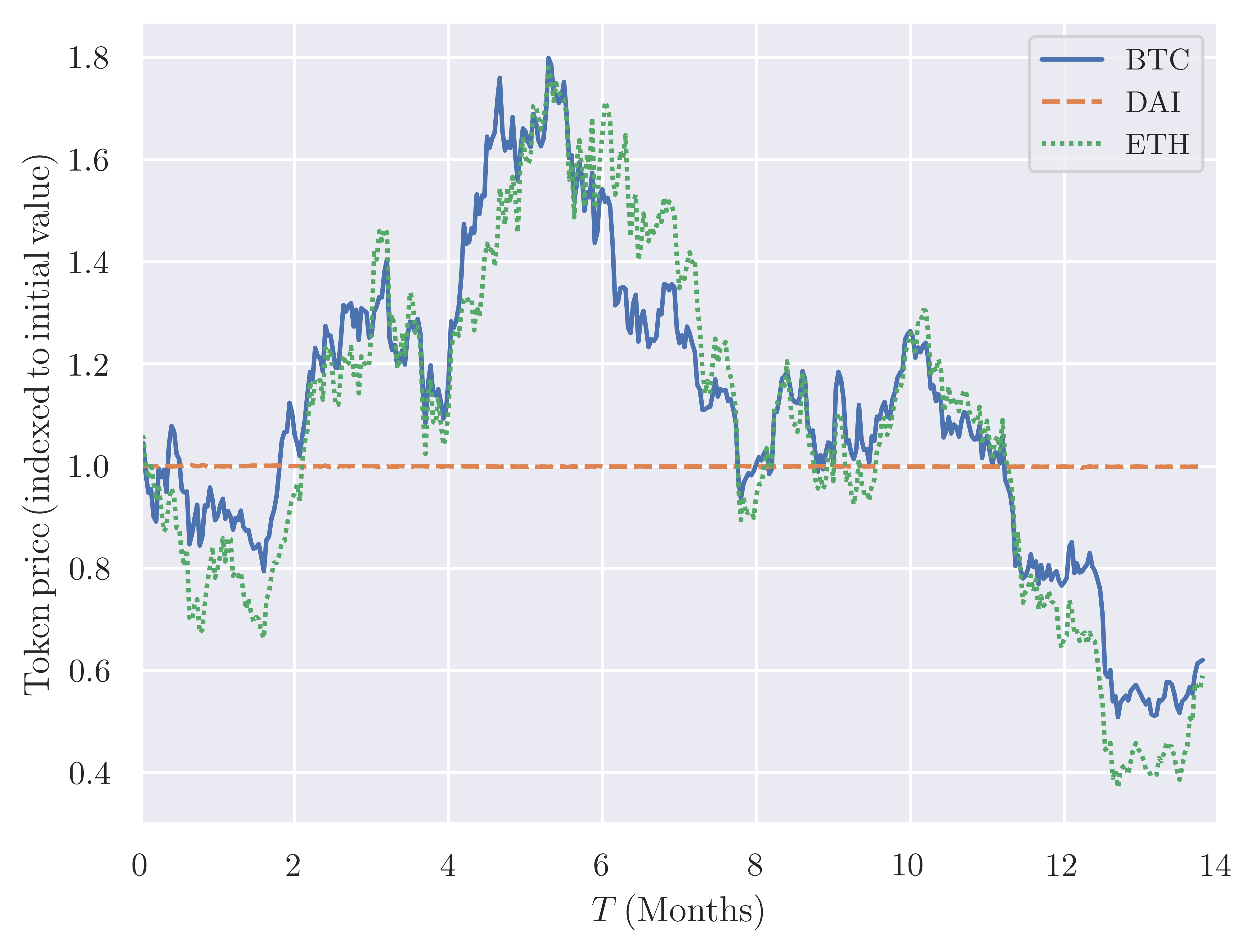}
    \caption{Asset prices for the basket over time (plotted indexed to initial prices).}
    \label{subfig:asset_prices}
     \end{subfigure}
 \hspace*{\fill}    
 \begin{subfigure}[b]{0.45\linewidth}
    \centering
         \centering
    \includegraphics[width=1\textwidth]{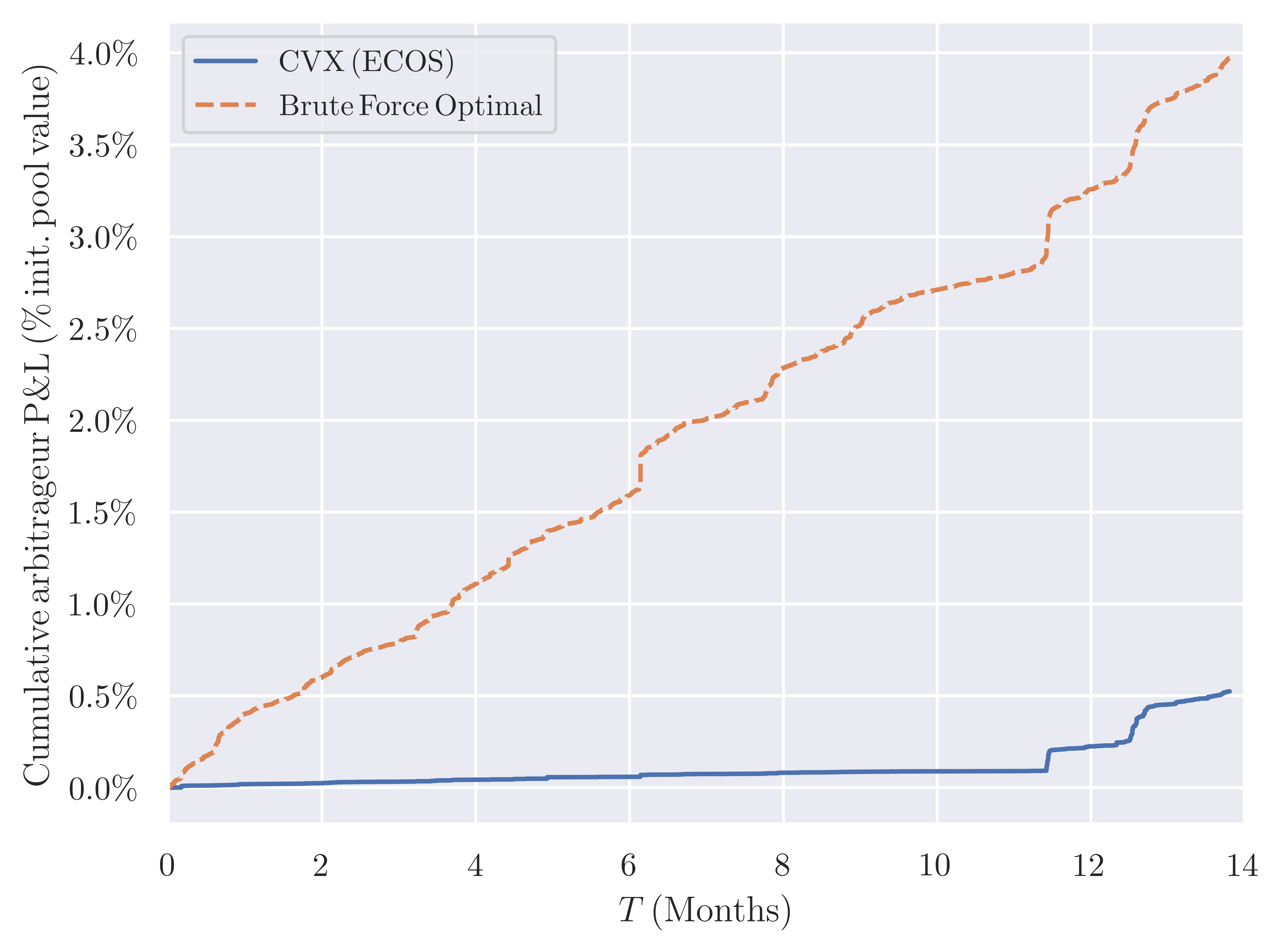}
    \caption{Cumulative arbitrageur profits as a fraction of initial pool value.}
    \label{subfig:cum_profits}
     \end{subfigure}
     \caption{Comparing arbitrage returns from competition between a simulated arbitrageur our approach and one using numerical solutions from convex optimisation. After every new market price comes in the convex optimiser arbitrageur gets priority (for free) in trading with the pool yet performs worse.}
     \label{fig:duelling_arbs}
\end{figure}

So in comparing multiple arbitrage algorithms over time we have to have them compete.
Here we compare our closed-form approach to convex optimisation (performed using CVXPY) on a historical backtest of a CFMM pool with $N=3$ run on a basket of ETH, BTC and DAI with uniform weights $w_i=\frac{1}{3}$ from June 2021 to July 2022.
$\gamma=0.003$ and initial pool value was 1m USD.
We plot the results in Figure~\ref{fig:duelling_arbs}.

To give the convex-solver-arbitrageur an edge over the closed form algorithm, the convex-solver-arb is provided the top transaction in the block every time, for free.
The closed form algorithm reaps the arbitrage opportunity a considerable number of times before convex-optimisation arbitrageur can.

\section {Conclusions}

We have described the closed-form expressions for the optimal arbitrage trade for $N$-token pools.
In a range of experiments we have demonstrated the speed and arbitrage performance advantages of this new closed-form approach over current approaches.

Coupling this with the ability to run the approach at even higher speeds on GPUs (due to its parallelisability), and to use take gradients through the process (enabling more advanced machine learning techniques for AMM strategy training~\cite{tfmm}) means that this approach may find use both for practitioners interested in performing arbitrage trades as well as those interested primarily in simulating \& modelling AMM pools.

There are on-chain implications too that are significant, given how this technique could be used not only for on-chain arbitrage bots that have the direct ability to structure the arbitrage trade (via access to the current state of the pool) in the same transaction that the trade is performed, but also for DEX order routers and DEX aggregator routers.
TSTORE advances, intelligent caching, and also signature detection heuristics being developed could each significantly reduce the computational cost of obtaining such a solution on-chain.

\bibliography{biblio}

\clearpage  %

\begin{appendices}
\renewcommand\thefigure{\thesection.\arabic{figure}}    
\renewcommand\theequation{\thesection.\arabic{equation}}   

\section{Derivation of multi-asset optimal arbitrage trade}
\label{app:basket_optimal}
We sub in Eq~\eqref{eq:ratio_rearranged} to Eq~\eqref{eq:tfmm_with_fees_phi} for all $j\neq i$, $i,j\in\mathcal{A}$:
\begin{align}
    1&=\left(1 + \frac{\gamma^{d_i} \Phi_i}{R_i}\right)^{\breve{w}_i} \prod_{\substack{j\neq i \\ j\in\mathcal{A}}} \left(\frac{\breve{w}_j \gamma^{d_j}}{\breve{w}_i \gamma^{d_i}}\frac{m_{p,i} R_i}{m_{p,j} R_j}\left(1 + \frac{\gamma^{d_i} \Phi_i}{R_i}\right)\right)^{\breve{w}_j}\\
    \Rightarrow 1&=\left(1 + \frac{\gamma^{d_i} \Phi_i}{R_i}\right) \prod_{\substack{j\neq i \\ j\in\mathcal{A}}} \left(\frac{m_{p,i} R_i}{\breve{w}_i \gamma^{d_i}}\right)^{\breve{w}_j} \prod_{\substack{j\neq i \\ j\in\mathcal{A}}} \left(\frac{\breve{w}_j \gamma^{d_j}}{m_{p,j} R_j}\right)^{\breve{w}_j}\\
    \Rightarrow 1&=\left(1 + \frac{\gamma^{d_i} \Phi_i}{R_i}\right)\left(\frac{m_{p,i} R_i}{\breve{w}_i \gamma^{d_i}}\right)^{1-\breve{w}_i}\prod_{\substack{j\neq i \\ j\in\mathcal{A}}} \left(\frac{\breve{w}_j \gamma^{d_j}}{m_{p,j} R_j}\right)^{\breve{w}_j}\\
    \Rightarrow \left(1 + \frac{\gamma^{d_i} \Phi_i}{R_i}\right)&=\left(\frac{\breve{w}_i \gamma^{d_i}}{m_{p,i} R_i}\right)^{1-\breve{w}_i}\prod_{\substack{j\neq i \\ j\in\mathcal{A}}} \left(\frac{m_{p,j} R_j}{\breve{w}_j \gamma^{d_j}}\right)^{\breve{w}_j}\\
    \Rightarrow \left(1 + \frac{\gamma^{d_i} \Phi_i}{R_i}\right)&=\left(\frac{\breve{w}_i \gamma^{d_i}}{m_{p,i} R_i}\right)^{1-\breve{w}_i}\prod_{\substack{j\neq i \\ j\in\mathcal{A}}} \left(\frac{m_{p,j} R_j}{\breve{w}_j \gamma^{d_j}}\right)^{\breve{w}_j}\\
    \Rightarrow \left(1 + \frac{\gamma^{d_i} \Phi_i}{R_i}\right)&=\left(\frac{\breve{w}_i \gamma^{d_i}}{m_{p,i} R_i}\right)^{1-\breve{w}_i}\frac{\breve{k}}{R_i^{\breve{w}_i}}\prod_{\substack{j\neq i \\ j\in\mathcal{A}}}\left(\frac{m_{p,j}}{\breve{w}_j \gamma^{d_j}}\right)^{\breve{w}_j}\\
    \Rightarrow \frac{\gamma^{d_i} \Phi_i}{R_i}&=\frac{\breve{k}}{R_i}\left(\frac{\breve{w}_i \gamma^{d_i}}{m_{p,i}}\right)^{1-\breve{w}_i}\prod_{\substack{j\neq i \\ j\in\mathcal{A}}} \left(\frac{m_{p,j}}{\breve{w}_j \gamma^{d_j}}\right)^{\breve{w}_j}-1\\
    \Rightarrow \forall i\in\mathcal{A},\, \Phi_i&=\gamma^{-d_i}\left(\breve{k}\left(\frac{\breve{w}_i \gamma^{d_i}}{m_{p,i}}\right)^{1-\breve{w}_i}\prod_{\substack{j\neq i \\ j\in\mathcal{A}}} \left(\frac{m_{p,j}}{\breve{w}_j \gamma^{d_j}}\right)^{\breve{w}_j}-R_i\right),
    \label{eq:app_optimal_trade}
\end{align}
where $\breve{k}:=\prod_{i\in\mathcal{A}}R_i^{\breve{w}_i}$.
There is also a slightly more symmetric form, which however is slightly less appropriate for implementation in a resource-constrained setting:
\begin{equation}
    \forall i\in\mathcal{A},\,\Phi_i=\gamma^{-d_i}R_i\left(\left(\frac{\breve{w}_i \gamma^{d_i}}{m_{p,i}R_i}\right)^{1-\breve{w}_i}\prod_{\substack{j\neq i \\ j\in\mathcal{A}}} \left(\frac{m_{p,j} R_j}{\breve{w}_j \gamma^{d_j}}\right)^{\breve{w}_j}-1\right)
    \label{eq:app_alt_optimal}
\end{equation}
\paragraph{Amplified Liquidity}
This approach also naturally extends to multi-token pools that make use of amplified liquidity \cite{tfmm_litepaper}.
There a pool's \emph{virtual reserves} are $\breve{R}_i=\nu R_i$ where $\nu$ is a (potentially varying from block to block) scalar and the trading invariant is that $\prod_{i} \left(\nu R_i + \gamma \Delta_i - \Lambda \right)^{w_i}= \nu \prod_{i} \left(R_i\right)^{w_i}$.
Applying the mapping $R_i\rightarrow\nu R_i$ to Eq~\eqref{eq:app_optimal_trade}, we get the optimal arbitrage trade for these pools:
\begin{equation}
    \forall i\in\mathcal{A},\, \Phi_i=\gamma^{-d_i}\left(\breve{\nu}\breve{k}\left(\frac{\breve{w}_i \gamma^{d_i}}{m_{p,i}}\right)^{1-\breve{w}_i}\prod_{\substack{j\neq i \\ j\in\mathcal{A}}} \left(\frac{m_{p,j}}{\breve{w}_j \gamma^{d_j}}\right)^{\breve{w}_j}-\nu R_i\right),
\end{equation}
where $\breve{\nu}:=\nu^{\sum_{i\in\mathcal{A}}\breve{w}_i}$, with the analogue of Eq~\eqref{eq:app_alt_optimal} similarly following.
\subsection{Heuristic for Trade Signature}
\label{app:heuristic}
A heuristic that works reasonably well for finding a workable trade signature $\v s$ in the close vicinity of the no-arb-region boundary is to look for, in effect, swap-level arb opportunities between each pair within the pool and use those to form $\v s$.  
Do as follows:
\paragraph{Find the zeros}
Construct the ratio between the (zero-fees) quoted prices of the pool (in the same num\'eraire as the market prices $\v m_p$) and the market prices: $\v \ell := V \frac{\v w}{\v R \v m_p}$ (where all operations are done elementwise) and $V=\sumiN m_{p,i}R_i$.
Then construct the `price quotient matrix' $\v\Gamma$, where each entry is $\Gamma_{i,j} = \ell_{i}/\ell_{j}$.
For any tokens $i$ for which $\sum_{j=1}^N \mathbb{I}_{\Gamma_{i,j}>\gamma^{-1}}=\sum_{j=1}^N \mathbb{I}_{\Gamma_{i,j}<\gamma}=0$, $s_i=0$.
\paragraph{Active trade direction}
For tokens $i$ with $s_i\neq0$, simply have $s_i=1$ if $\ell_i>1$ and $s_i=-1$ if $\ell_i<1$.
\section{Reduced form of Trading Function for G3Ms}
\label{app:reduced}
Starting with Eq~\eqref{eq:tfmm_with_fees} and making the substitutions from $\{\v\Delta,\v\Lambda\}$ to $\{\v\Phi,\v d\}$:
\begin{equation}
    \prod_{i=1}^N \left(R_i + \gamma^{d_i} \Phi_i\right)^{w_i} \geq k.
\end{equation}
The most profitable trade will be when equality is reached.
Divide by $ \prod_{i=1}^N\left(R_i\right)^{w_i}=k$ to obtain
\begin{align}
     \prod_{i=1}^N \frac{\left(R_i + \gamma^{d_i} \Phi_i\right)^{w_i}}{\left(R_i\right)^{w_i}} &= 1\\
     \prod_{i=1}^N \left(1+ \frac{\gamma^{d_i} \Phi_i}{R_i}\right)^{w_i} &= 1.
\end{align}
As $\Phi_i=0$ for $i\notin\mathcal{A}$, $\left(1+ \frac{\gamma^{d_i} \Phi_i}{R_i}\right)=1$ for $i\notin\mathcal{A}$, so in the product we can consider only the set of indices $\mathcal{A}$, giving us
\begin{equation}
      \Rightarrow\prod_{i\in\mathcal{A}} \left(1+ \frac{\gamma^{d_i} \Phi_i}{R_i}\right)^{w_i} = 1.
\end{equation}
We can now take the $\left(\sum_{j\in\mathcal{A}}w_j\right)^{\mathrm{th}}$ root, so we obtain
\begin{align}
      \Rightarrow\prod_{i\in\mathcal{A}} \left(1+ \frac{\gamma^{d_i} \Phi_i}{R_i}\right)^\frac{w_i}{\sum_{j\in\mathcal{A}}w_j} &= 1\\
      \Rightarrow \prod_{i\in\mathcal{A}} \left(1+ \frac{\gamma^{d_i} \Phi_i}{R_i}\right)^{\breve{w}_i} &= 1
\end{align}
as required.
\newpage

\end{appendices}

\end{document}